\documentclass[twoside,fleqn]{article}
\usepackage[dvips]{epsfig}
\usepackage{espcrc2}

\newcommand{\MeV}{\,\rm{MeV}}

\title{Lattice Study of the $H$ Dibaryon}
\author{J.~W.~Negele, A.~Pochinsky and B.~Scarlet\address{Center for
                Theoretical Physics, Massachusetts Institute of Technology,
                Cambridge, MA 02139, USA}\thanks{Based on the talk presented
                by A.V.P.  Work supported by the U.S. Department of Energy 
               (DOE) under cooperative research agreement DE-FC02-94ER40818.}}

\begin{document}

\begin{abstract}
The mass of the lowest spin-zero, strangeness-$(-2)$ flavor singlet state
in the dibaryon sector has been calculated in quenched QCD on
$16^3\times32$ and $24^3\times32$ lattices at $\beta=5.85$ to study
whether the energy of the proposed $H$ dibaryon is near or below the
$\Lambda\Lambda$ threshold. Preliminary results indicate that finite
lattice volume artifacts overestimate the binding, and that on the
largest lattice $m_H$ is of the order of $100\MeV$ above the 
$\Lambda\Lambda$ threshold.
\end{abstract}

\maketitle 

\section{Introduction}

Based on physical bag model arguments and the magnetic hyperfine
interaction, Jaffe suggested that the lowest bound state in the
dibaryon sector would be a spin $0$ strangeness $-2$ $SU(3)$ flavor
singlet, and that the energy of this $H$ dibaryon could be near or
below the $\Lambda\Lambda$ threshold~\cite{Jaffe}.  Despite two
decades of effort, experiments have neither found it nor excluded the
possibility of exotic multiquark systems ($Q^n\bar{Q}^m$, for
$n+m>3$).

If a six quark system exists as a bound state, a single gluon exchange
will contribute
\begin{equation}
\Delta = -\sum_{ij} (\vec{\sigma}_i \cdot \vec{\sigma}_j)
                   (\vec{\lambda}_i \cdot \vec{\lambda}_j) M_{ij}
\end{equation}
to the mass \cite{Jaffe} yielding Jaffe's original bag model estimate
of $m_H=2150\MeV$, $81\MeV$ below the $2231\MeV$ $\Lambda\Lambda$
threshold. Subsequent bag model predictions \cite{bags} yield the mass
range $1.03-2.3$ GeV. Although early lattice calculations discussed
below were not definitive \cite{Mac,Iwasaki}, the success of
contemporary spectroscopy in quenched QCD motivated us to attempt a
new lattice calculation of the $H$ dibaryon.

\section{Lattice Parameters}
We have studied two lattice sizes, $16^3\times32$ and $24^3\times32$
at quenched $\beta=5.85$, corresponding to a lattice spacing $a\approx
0.13(3)\,\rm{fm}$. Wilson fermion propagators were calculated at seven
values of the hopping parameter for the bare quark mass in the range
$30-300\MeV$ using a point source at $t=0$. For the fermions the
boundary conditions were periodic in spatial directions and hard wall
(Neumann) in the temporal direction. This choice allowed us to obtain
a signal at large euclidian times which is crucial for the
calculation. A total of 56 configurations at $16^3\times32$ and 40 at
$24^3\times32$ were used.

Once the quark propagators were calculated, the two-point functions
for the $H$ as well as for assorted mesons ($\pi$, $\rho$, $\phi$, $K$
and $K^{*}$) and hadrons ($N$, $\Lambda$ and $\Xi$) were constructed
with the three-momentum $\vec{p}=0$ at the sink. The lattice spacing
is defined by $n_N$, the degenerate $u$ and $d$ quark masses are
defined by $m_\pi/m_N$ and the strange quark mass is defined by
$m_\Lambda/m_N$. These definition were chosen to absorb as many of the
errors arising from the quenched approximation as possible.

\section{\protect\boldmath$H$ Two-point Function}
While the $H$ propagator could be expressed in terms of two-hadron
states, this form is not computationally feasible due to numerous
duplicate terms and mutual cancellations.  Instead, we constructed the
polynomial arising from contracting $J=0$, $S=-2$, $B=2$ sources and
sinks of the form $\phi_H=\sum c_i \epsilon^{abc}\epsilon^{a'b'c'}
\epsilon_{\alpha\beta} \epsilon_{\gamma\mu} \epsilon_{\nu\lambda}
(uuddss)^{\alpha\beta\gamma\mu\nu\lambda}_{abca'b'c'}$, and
manipulated it analytically to reduce the computational complexity by
a factor of 200. The final expression was calculated by two independent
methods and symmetries of the result were checked.

As a result, the machine time was divided evenly between the Dirac
inverter and two-point function calculations for 16 combinations of
heavy and light quark masses.

\section{Measurements}
In the absence of finite boundary effects, one may identify the lowest mass
by writing the zero momentum two point function
\begin{eqnarray}
D(t)&=&\int d^3x\, e^{ipx}D(0;x,t)|_{p=0} \nonumber\\
    &=&\sum_{|n\rangle} a_n e^{-\langle n|\hat H|n\rangle t}
\end{eqnarray}
where $n$ runs over all states with given quantum numbers and
extracting the exponent with smallest mass. In our case, there is an
additional correction arising from the hard wall, so we approximate $D(t)$ in
the region of interest by three terms:
\begin{equation}\label{3-mass}
D(t)\approx a_1 e^{-m_1t} + a_2 e^{-m_2t}
      + a_3 e^{-m_3t},
\end{equation}
and fit data in the window $ t_1 \le t \le t_2$. In this case, $m_1$
represents contributions from the excited states, $m_2$ is the ground
state mass and $m_3$ accounts for the wall effects. This fit has been
chosen over a simpler one mass fit because it is less sensitive to
the window size and position. In addition, the resulting values of $m_2$ are
less contaminated by higher excitations and wall effects.
\begin{figure}[t!]
\begin{center}
\epsfig{file=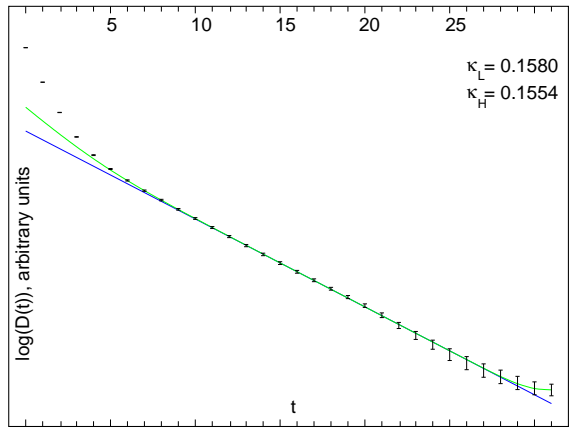,width=7.5cm}
%\vbox to 1in{\vfill lambda.eps \vfill}
\end{center}
\vspace{-30pt}
\caption{Lambda Propagator}
\label{L-plot}
\vspace{-20pt}
\end{figure}
\begin{figure}[t!]
\begin{center}
\epsfig{file=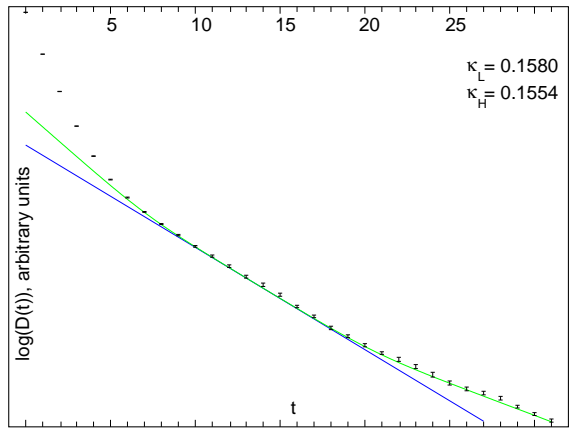,width=7.5cm}
%\vbox to 1in{\vfill H.eps \vfill}
\end{center}
\vspace{-30pt}
\caption{Dibaryon Propagator}
\label{H-plot}
\vspace{-20pt}
\end{figure}
 Typical results of the fit to the $\Lambda$ and $H$ on the
$24^3\times32$ lattice are shown on figures~\ref{L-plot}
and~\ref{H-plot} respectively. As one can see, the wall effects on
baryons are rather small, while, as expected from the much more rapid
fall off for the dibaryon, there is a larger effect on the $H$. We note
that in all cases, eq.~(\ref{3-mass}) yielded a good fit to the linear
interior region that was stable against reasonable changes in the
$(t_1,t_2)$ window. Errors in the mass $m_2$ were determined by
least squares analysis. Physical masses were determined by linear
extrapolation in the light and strange quark masses.

\section{Baryon Spectrum}
Since $m_u$ and $m_d$ are fixed by the nucleon mass, in the
non-strange sector, $m_\rho$ is a test both for finite size effects
and quenching errors. On both lattice sizes we obtained comparable results,
and, as expected~\cite{Aoki}, $m_\rho$ is about $10\%$ lighter than the
experimental value.

In the strange sector, having set the strange quark mass by
$m_\Lambda$, the lattice results [and corresponding experimental values]
are: $m_\phi=990(6)\MeV$ [$1020\MeV$], $m_K= 600(5)\MeV$ [$(m_{K^+}
+m_{K^0})/2=496\MeV$], $m_{K^*}= 840(20)\MeV$ [$894\MeV$] and
$m_\Xi=1360(34)\MeV$ [$(m_{\Xi^0} +m_{\Xi^-})/2=1318\MeV$]. The errors
are comparable with other quenched calculations.

\section{\protect\boldmath$H$ mass}
For the lattice sizes we are able to treat, $m_H$ has strong finite
lattice size effects.

On the $16^3\times32$ lattice, the extrapolation to the physical $m_u$
and $m_s$ gives $m_H=1950(60)\MeV$, just above the deuteron
threshold. However, the $2\,\rm{fm}$ lattice appears too small to rule
out considerable finite size effects.

Results from the larger lattice, $24^3\times32$ at the same value of
$\beta$ yield an unbound $H$ at $2340(20)\MeV$. The dibaryon
remain heavier than two $\Lambda$'s at all combinations of the hopping
parameters, as shown in table~\ref{large-lattice}.
\begin{table}[t!]
\caption{Masses of the $H$ and $\Lambda$ in $24^3\times32$ lattice.}
\label{large-lattice}
\setlength{\tabcolsep}{5pt}
\newlength{\digitwidth} \settowidth{\digitwidth}{\rm 0}
\catcode`?=\active \def?{\kern\digitwidth}
\label{tab:xxx}
\begin{tabular*}{7.5cm}{r@{\extracolsep{10pt}}
                        r@{\extracolsep{30pt}}
                        r@{\extracolsep{20pt}}r}
\hline
      &           & \multicolumn{2}{c}{$ma^{-1}$} \\
\cline{3-4}
 \multicolumn{1}{c}{$\kappa_H$} & \multicolumn{1}{c}{$\kappa_L$} &
 \multicolumn{1}{c}{$H$} & \multicolumn{1}{c}{$\Lambda$} \\
\hline
0.1520 & 0.1568 &  2.091(09) &  1.0143(06)  \\
       & 0.1580 &  1.988(10) &  0.9602(08)  \\
       & 0.1592 &  1.858(14) &  0.9053(17)  \\
       & 0.1601 &  1.754(20) &  0.8586(24)  \\
                                         &&&\\
0.1539 & 0.1568 &  2.021(08) &  0.9790(07)  \\
       & 0.1580 &  1.930(10) &  0.9249(10)  \\
       & 0.1592 &  1.799(15) &  0.8681(20)  \\
       & 0.1601 &  1.680(22) &  0.8209(23)  \\
                                         &&&\\
0.1554 & 0.1568 &  1.968(07) &  0.9501(08)  \\
       & 0.1580 &  1.878(11) &  0.8950(12)  \\
       & 0.1592 &  1.753(17) &  0.8371(23)  \\
       & 0.1601 &  1.642(26) &  0.7876(23)  \\
                                         &&&\\
0.1568 & 0.1568 &  1.916(07) &  0.9221(10) \\
       & 0.1580 &  1.823(12) &  0.8647(15) \\
       & 0.1592 &  1.705(22) &  0.8060(26) \\
       & 0.1601 &  1.605(33) &  0.7559(25) \\
\hline
\end{tabular*}
\end{table}

The sign and magnitude of the finite volume dependence indicate that
interactions with periodic images are attractive, and significant at a
separation of $2\,\rm{fm}$. Since the attraction decreases with
increasing volume and the $H$ is already unbound by the order of
$110\MeV$ on $24^3\times32$ lattice, we believe the infinite volume
extrapolation of quenched QCD is even more unbound.

\section{Comparison with Previous Work}
There were two previous attempts to calculate $m_H$ on the lattice a
decade ago. Mackenzie and Tacker \cite{Mac} did not see a bound
dibaryon on a rather small lattice $6\times6\times12\times18$. Three
years later Iwasaki, Yoshi\'e and Tsuboi \cite{Iwasaki} used a larger
lattice ($16^4\times48)$ and obtained a very strongly bound $H$,
between $1450(250)\MeV$ and $1710(140)\MeV$, depending on the
extrapolation. Since they used an unconventional gauge action and
quark masses considerably heavier than those used in our calculation,
direct comparison of numerical results is difficult. However, their
result is consistent with the fact that interactions with images tend
to overbind.

\section{Conclusion and Future Plans}
To strengthen our preliminary conclusion that the $H$ is unbound in
quenched QCD, we intend to recalculate the small lattice propagators
at the same hopping parameters at the large lattice, so that more
direct study of the finite size effects can be made. We also will
calculate the two-point functions for dibaryons with other quantum
numbers. In addition, we will extrapolate hadron masses in $m_\pi$ and
$m_K$ rather than in the quark masses.

\section*{Acknowledgment}
The authors are grateful to Robert Jaffe and Sergei Bashinsky for fruitful
discussions, and to Sun Microsystems for providing the 24
Gflops E5000 SMP cluster and the Wildfire prototype system on which
our calculations were performed.

\end{document}